\documentclass[a4paper,fleqn,usenatbib]{mnras}

\usepackage[T1]{fontenc}
\usepackage{ae,aecompl}
\usepackage{threeparttable}
\usepackage{arydshln}

\usepackage{graphicx}	
\usepackage{amsmath}	
\usepackage{amssymb}	

\title[Diffuse H{\small \,II} regions of the gas streams around Sgr A*]{Study of diffuse H{\small \,II} regions potentially forming part of the gas streams around Sgr A*}
\author[J. Armijos-Abenda\~no]{J. Armijos-Abenda\~no$^1$\thanks{E-mail: jairo.armijos@epn.edu.ec}, E. L\'opez$^1$, J. Mart\'in-Pintado$^2$, 
A. B\'aez-Rubio$^3$, M. Aravena$^4$,
\newauthor M. A. Requena-Torres$^5$, S. Mart\'in$^{6,7}$, M. Llerena$^1$, F. Ald\'as$^1$, C. Logan$^8$
\newauthor and A. Rodr\'iguez-Franco$^9$
\\
$^1$Observatorio Astron\'omico de Quito, Escuela Polit\'ecnica Nacional, Av. Gran Colombia S/N, Interior del Parque La Alameda, 170136,\\
Quito, Ecuador\\
$^2$Centro de Astrobiolog\'ia (CSIC, INTA), Ctra a Ajalvir, km 4, 28850, Torrej\'on de Ardoz, Madrid, Spain\\
$^3$Instituto de Astronom\'ia, Universidad Nacional Aut\'onoma de M\'exico, Circuito de la Investigaci\'on Cient\'ifica s/n, Ciudad Universitaria,\\
Del. Coyoac\'an, 04510, CDMX, Mexico\\
$^4$N\'ucleo de Astronom\'ia, Facultad de Ingenier\'ia y Ciencias, Universidad Diego Portales, Av. Ej\'ercito 441, Santiago, Chile\\
$^5$Department of Astronomy, University of Maryland, College Park, MD 20742, USA\\
$^6$European Southern Observatory, Alonso de C\'ordova 3107, Vitacura, Santiago, Chile\\
$^7$Joint ALMA Observatory, Alonso de C\'ordova 3107, Vitacura, Santiago, Chile\\
$^8$H. H. Wills Physics Laboratory, University of Bristol, Tyndall Ave., Bristol, BS8 1TL, United Kingdom\\
$^9$Facultad de \'Optica y Optometr\'ia, Departamento de Matem\'atica Aplicada (Biomatem\'atica), Universidad Complutense de Madrid, Avenida\\
de Arcos de Jal\'on, 118, E-28037 Madrid, Spain
}

\date{Accepted XXX. Received YYY; in original form ZZZ}

\pubyear{2015}

\begin{document}
\label{firstpage}
\pagerange{\pageref{firstpage}--\pageref{lastpage}}
\maketitle

\begin{abstract}
We present a study of diffuse extended ionised gas toward three clouds located in the Galactic Centre (GC). One line of sight (\emph{LOS}) is toward the \mbox{20 km s$^{-1}$} cloud (\emph{LOS}$-$0.11) in the Sgr A region, another \emph{LOS} is toward the \mbox{50 km s$^{-1}$} cloud (\emph{LOS}$-$0.02), also in Sgr A, while the third is toward the Sgr B2 cloud (\emph{LOS}+0.693). The emission from the ionised gas is detected from H$n\alpha$ and H$m\beta$ radio recombination lines (RRLs). He$n\alpha$ and He$m\beta$ RRL emission is detected with the same $n$ and $m$ as those from the hydrogen RRLs only toward \emph{LOS}+0.693. RRLs probe gas with positive and negative velocities toward the two Sgr A sources. The H$m\beta$ to H$n\alpha$ ratios reveal that the ionised gas is emitted under local thermodynamic equilibrium conditions in these regions. We find a He to H mass fraction of 0.29$\pm$0.01 consistent with the typical GC value, supporting the idea that massive stars have increased the He abundance compared to its primordial value. Physical properties are derived for the studied sources. We propose that the negative velocity component of both Sgr A sources is part of gas streams considered previously to model the GC cloud kinematics. Associated massive stars with what are presumably the closest H{\small \,II} regions to \emph{LOS}$-$0.11 (positive velocity gas), \emph{LOS}$-$0.02 and \emph{LOS}+0.693 could be the main sources of UV photons ionising the gas. The negative velocity components of both Sgr A sources might be ionised by the same massive stars, but only if they are in the same gas stream.
\end{abstract}




\section{Introduction}
\begin{figure}
\includegraphics[width=70mm]{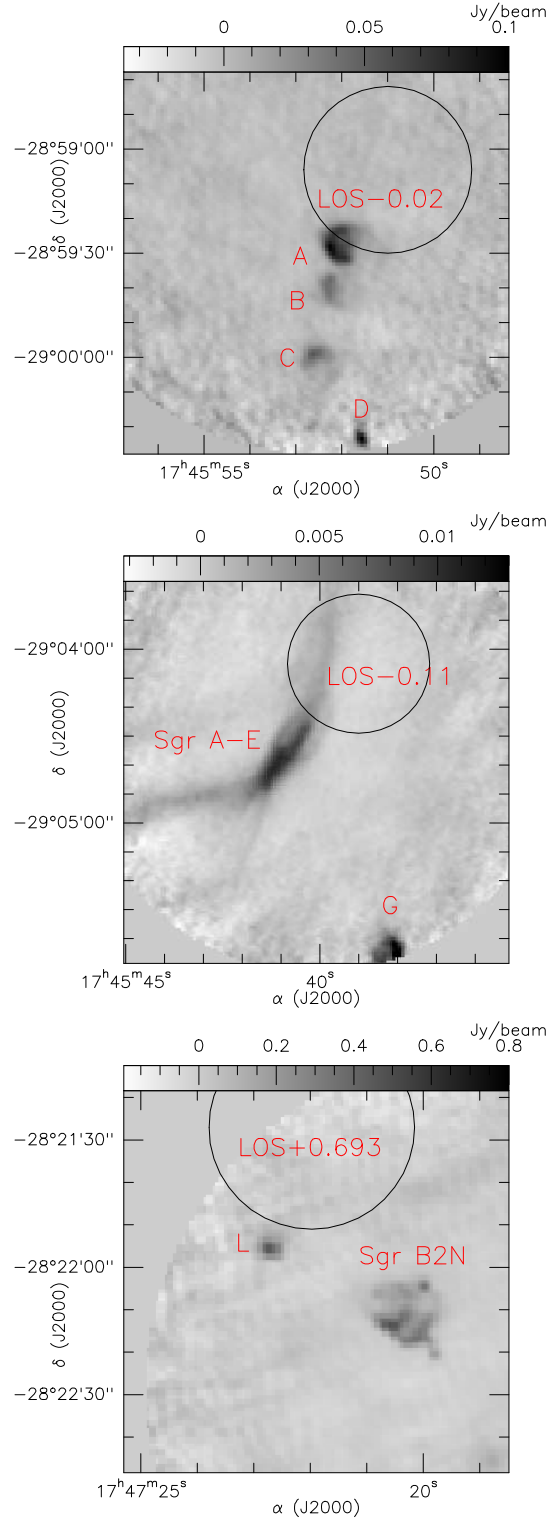}
\caption{VLA radio$-$continuum maps at 24.5 GHz toward the three \emph{LOSs} observed by us (see Section \ref{subsection_of_archival_VLA_data}). The three \emph{LOSs} are shown as black circles with the size of the GBT beam of 48 arcsec at 13.09 GHz. \textbf{Upper panel:} The region A partly falls inside \emph{LOS}$-$0.02. The regions B, C and D are also seen in the field. \textbf{Middle panel:} \emph{LOS}$-$0.11 overlaps with part of the non$-$thermal source Sgr A$-$E \citep{Lu03}. The region G appears to be the closest H{\small\,II} region to \emph{LOS}$-$0.11 \citep{Ho85}. \textbf{Bottom panel:} \emph{LOS}+0.693 lies close to the H{\small\,II} region L located northeast of \mbox{Sgr B2N}.}\label{fig1}
\end{figure}

The proximity of the Galactic Centre (GC), at a distance of about \mbox{$7.86$ \textrm{kpc}} \citep{Boehle16}, offers a unique opportunity to look at a galactic nucleus in great detail.
Several studies have been carried out to establish the physical properties and the kinematics of ionised gas toward the main compact H{\small \,II} regions located at the centre of the Galaxy \citep{Ho85,Mehringer93,Zhao93,Mills11}.
Sgr A West, located around the supermassive black hole Sgr A*, is a spiral$-$shaped region of ionised gas whose emission is thermal in nature \citep{Ekers83}. Sgr A East is a non$-$thermal source surrounding Sgr A West in projection \citep{Ekers83}. There is also a group of four H{\small \,II} regions, known collectively as G$-$0.02$-$0.07, made up of the regions denoted as A, B, C, and D (see Fig.~\ref{fig1}, upper panel). G$-$0.02$-$0.07 is located at a projected distance of \mbox{$\sim$6 \textrm{pc}} from Sgr A*.
These H{\small \,II} regions likely reside within the \mbox{50 km s$^{-1}$} cloud\footnote{The name is given by its local standard of rest radial velocities.} \citep{Goss85,Mills11}, one of the massive clouds in the Sgr A complex. 
Sgr A East may be impacting the 50 km s$^{-1}$ cloud at its west side \mbox{\citep{Serabyn92}.} Massive O stars are thought to be ionising the A$-$D regions \citep{Lau14}. Using line to continuum ratios, \cite{Goss85} found electron temperatures in the range of $\sim$5000$-$7000 K for the four compact H{\small\,II} regions.

Another H{\small \,II} region labelled as G (see Fig.~\ref{fig1}, \mbox{middle} panel), located at \mbox{$\sim$13 \textrm{pc}} in projection from Sgr A*, is thought to be excited by one O9 or five B0 stars \citep{Ho85}. The region G appears to be embedded in the \mbox{20 km s$^{-1}$} cloud \citep{Armstrong89}, another massive cloud in the Sgr A complex. Sgr A$-$E is considered a non$-$thermal source \citep{Lu03}, which lies close to the region G (see Fig.~\ref{fig1}). \cite{Armstrong89} found an electron temperature of $\sim$7500 K for the region G.

\citet{Zhao93} studied five H{\small \,II} regions (identified as H1 through to H5) located between Sgr A West and the Arched Filaments H{\small \,II} complex containing a group of curved ridges showing velocities from 15 to -70 km s$^{-1}$ \citep{Lang01}. The H1$-$H5 sources show gas velocities from \mbox{-20} to \mbox{-60 km s$^{-1}$,} which seem to be associated with a \mbox{-30 km s$^{-1}$} cloud \citep{Zhao93}. However, negative velocities of the ionised gas are not only observed toward the H1$-$H5 regions and the Arched Filaments H{\small \,II} complex, as previously thought, but also toward many other regions of the \mbox{Sgr A} complex. In fact, a GC large-scale map obtained by \cite{Royster14} shows ionised gas toward the Sgr A complex with negative velocities reaching up to \mbox{$\sim$-130 km s$^{-1}$}. A recent position-velocity map of the C{\small \,II} emission \citep{Langer17}, which is considered as a good tracer of the ionised gas, shows a similar distribution as in the map obtained by \cite{Royster14}. Clouds of diffuse ionised gas in Sgr A with velocities from $\sim$-130 to +130 km s$^{-1}$ are shown on the channel maps of the C{\small \,II} emission obtained by \mbox{\cite{Garcia15}}.

On the other hand, the Sgr B2 complex lies at a projected distance of $\sim$120 pc from the GC. This complex contains many dozens of compact and ultracompact H{\small \,II} regions \citep{Gaume95,DePree05}. Many of these H{\small \,II} regions are associated with the Sgr B2 north (N), main (M) and south (S) hot cores where star formation is taking place \citep{Gordon93}.
The ionised gas in the Sgr B2 complex shows velocities predominantly in the range of 50$-$70 \mbox{km s$^{-1}$} \citep{Mehringer93}. There is a H{\small \,II} region labelled as L \citep{Mehringer93} that is located at a projected distance of $\sim$1.6 \textrm{pc} from Sgr B2N (see Fig.~\ref{fig1}, bottom panel). 
The region L has an electron temperature of $\sim$6500 K \citep{Mehringer93} and it is believed to be excited by one O5.5 star \citep{Gaume95}.

The 20 and 50 km s$^{-1}$ clouds are considered as part of a set of clouds moving on stable x$_2$ orbits around the GC in a 100$\times$60 \textrm{pc} elliptical and twisted ring \citep{Molinari11}.
In this scenario both clouds are located in the front region of the ring while its background gas, which is around both clouds as seen in projection, show
velocities from $\sim$0 to \mbox{-60 km s$^{-1}$} \citep{Molinari11}.
\cite{Kruijssen15} also modelled the gas kinematics studied by \cite{Molinari11}, reproducing the kinematics of molecular gas using an open gas stream divided into four gas streams orbiting the GC. The back side of the open stream is composed of streams 3 and 4, while streams 1 and 2 are two ends of the open stream located at its front side \citep{Kruijssen15}. The 20 and 50 km s$^{-1}$ clouds are contained in the gas stream 1. A recent study \citep{Langer17} revealed that the ionised gas velocities of the Sgr A and Sgr B2 clouds are better explained by the gas streams proposed by \cite{Kruijssen15} rather than by the elliptical ring proposed by \cite{Molinari11}. \cite{Henshaw16} found that two spiral arms or gas streams reproduce the molecular gas distribution of several GC clouds. Since no known physical model explains the spiral arms \citep{Henshaw16}, open streams might be the most likely structure.

In this paper we focus on studying the physical properties and kinematics of the diffuse ionised gas of selected GC regions.
Using radio recombination lines (RRLs) observed with the Green Bank Telescope (GBT) of NRAO\footnote{The National Radio Astronomy Observatory is a facility of the National Science
Foundation, operated under a cooperative agreement by Associated Universities, Inc.}, we find that RRLs show positive and negative velocities toward two
lines of sight (\emph{LOS}) in the Sgr A complex, one toward the \mbox{50 km s$^{-1}$} cloud (\emph{LOS}$-$0.02) and another toward the \mbox{20 km s$^{-1}$} cloud (\emph{LOS}$-$0.11). We also study the ionised gas along one \emph{LOS} in the Sgr B2 complex (\emph{LOS}+0.693) for comparison purposes.
Fig.~\ref{fig1} shows the observed positions of the three \emph{LOS}, where other GC sources are indicated.
As indicated in Fig.~\ref{fig1} \emph{LOS}$-$0.02 covers part of the emission arising from the H{\small \,II} region A. The region G appears to be the closest thermal H{\small \,II} region to \emph{LOS}$-$0.11 \citep{Ho85} since Sgr A$-$E is considered a non$-$thermal source in nature \citep{Lu03,Yusef05}. \emph{LOS}+0.693 lies close to the H{\small \,II} region L (see Fig.~\ref{fig1}, bottom panel).

This paper is organized as follows. In Section \ref{Observation} we present the observations and data used in this work. We present the main results in Section \ref{Results},
focusing on the line identification of RRLs 
and Gaussian fits in Section \ref{Gaussian_fits}, the local thermodynamic equilibrium of the ionised gas in Section \ref{LTE_conditions}, helium to hydrogen ratio in Section \ref{He_to_H_ratio}, and electron densities and the number of Lyman continuum photons in Section \ref{Physical_prop}. 
We discuss whether the RRL emission detected with the GBT is extended and diffuse in Section \ref{Extended_emission}, the kinematics of the ionised gas in Section \ref{Kinematics}, and the sources of gas ionisation in Section \ref{Ionization}. Finally, the conclusions of this work are presented in Section \ref{Conclusions}.


\section{Observations and data reduction}\label{Observation}

The observations were carried out with the NRAO 100$-$m Green Bank Telescope (GBT) in July$-$October 2009. We used
the Ku$-$band receiver connected to the spectrometer that provided four \mbox{200 MHz} spectral windows in two polarizations. This
configuration provides a spectral resolution of \mbox{24.4 kHz} or \mbox{0.6 km s$^{-1}$.} Spectra were calibrated using a noise tube and the line intensities, affected by 20 per cent uncertainties, are given in T$_\mathrm{A}^*$ scale.
The position-switched mode was used during the observations. As mentioned, the studied \emph{LOS}s are shown in Fig.~\ref{fig1}.
The angular resolution is 45 arcsec at 14.19 GHz, which corresponds to $\sim$1.7 \textrm{pc} at the distance of the GC. We used the reference positions selected and verified by \cite{Martin08}, which were originally based on large scale CS maps \citep{Bally87}. 
The three observed \emph{LOS} positions and their reference positions are indicated in Table \ref{tab_obs}.

Using the {\small GBTIDL} package\footnote{{\small GBTIDL} is an NRAO data reduction package, written in the IDL language for the reduction of GBT data.}, we inspected all scans of the
SDFITS files, and the baseline subtraction and average were applied to the calibrated spectra. 
Then the data were imported into the {\small MADCUBA} package\footnote{This package have been developed at the Centro de Astrobiolog\'ia. More information about this package in 
http://cab.inta-csic.es/madcuba/Portada.html.} for further processing. The spectra were smoothed to a velocity resolution of \mbox{$\sim$5 km s$^{-1}$} appropriate for the RRL widths, $\Delta v_\mathrm{r}$, of \mbox{$\sim$30 km s$^{-1}$} observed in the GC \citep{Mehringer93}.

\begin{table}
\scriptsize
\caption{Observed positions and their references}\label{tab_obs}
\begin{tabular}{ccccc}
\hline
Source & \multicolumn{2}{c}{Position} & \multicolumn{2}{c}{Reference} \\
       &  RA(J2000) & DEC(J2000) & RA(J2000) & DEC(J2000) \\
\hline
\emph{LOS}$-$0.02  & $17^{\rmn{h}}45^{\rmn{m}}51.0^{\rmn{s}}$ & $-28\degr59\arcmin06.0\arcsec$ & $17^{\rmn{h}}46^{\rmn{m}}00.1^{\rmn{s}}$ & $-29\degr16\arcmin47.2\arcsec$ \\
\emph{LOS}$-$0.11  & $17^{\rmn{h}}45^{\rmn{m}}39.0^{\rmn{s}}$ & $-29\degr04\arcmin05.0\arcsec$ & $17^{\rmn{h}}46^{\rmn{m}}00.1^{\rmn{s}}$ & $-29\degr16\arcmin47.2\arcsec$ \\
\emph{LOS}+0.693   & $17^{\rmn{h}}47^{\rmn{m}}22.0^{\rmn{s}}$ & $-28\degr21\arcmin27.0\arcsec$ & $17^{\rmn{h}} 46^{\rmn{m}} 23.0^{\rmn{s}}$ & $-28\degr 16\arcmin 37.3\arcsec$ \\
\hline
\end{tabular}
\end{table}
\subsection{Archival VLA data}
\label{subsection_of_archival_VLA_data}
To find out whether the emission detected with the GBT is affected by emission arising from compact H{\small \,II} regions (see discussion in Section \ref{Extended_emission}), we have used VLA data at 24.5 GHz available in the NRAO archive\footnote{https://archive.nrao.edu/}. The VLA data reduction and imaging were done using the \textsc{CASA} package\footnote{http://casa.nrao.edu/} (version 4.7.0). The observations were carried out in 2012 using the DnC configuration. We have build continuum maps, shown in Fig.~\ref{fig1}, and also a H64$\alpha$ cube for the \emph{LOS}$-$0.02 region as this information will be required in Section \ref{Extended_emission}. The continuum maps and cube were obtained using the \texttt{clean} task of \textsc{CASA}.
The spatial resolution of the maps and cube is 2.52$\times$2.47 arcsec$^2$. 
The cube has a rms noise of $\sim$4 mJy beam$^{-1}$ per channel, while the continuum maps of both Sgr A regions and the \emph{LOS}+0.693 region have rms noises of $\sim$2 and $\sim$20 mJy beam$^{-1}$, respectively.

\section{Results}\label{Results}
\subsection{Line identification and Gaussian fits}\label{Gaussian_fits}

\begin{figure}
\includegraphics[trim=0 150 0 0, clip, width=\columnwidth]{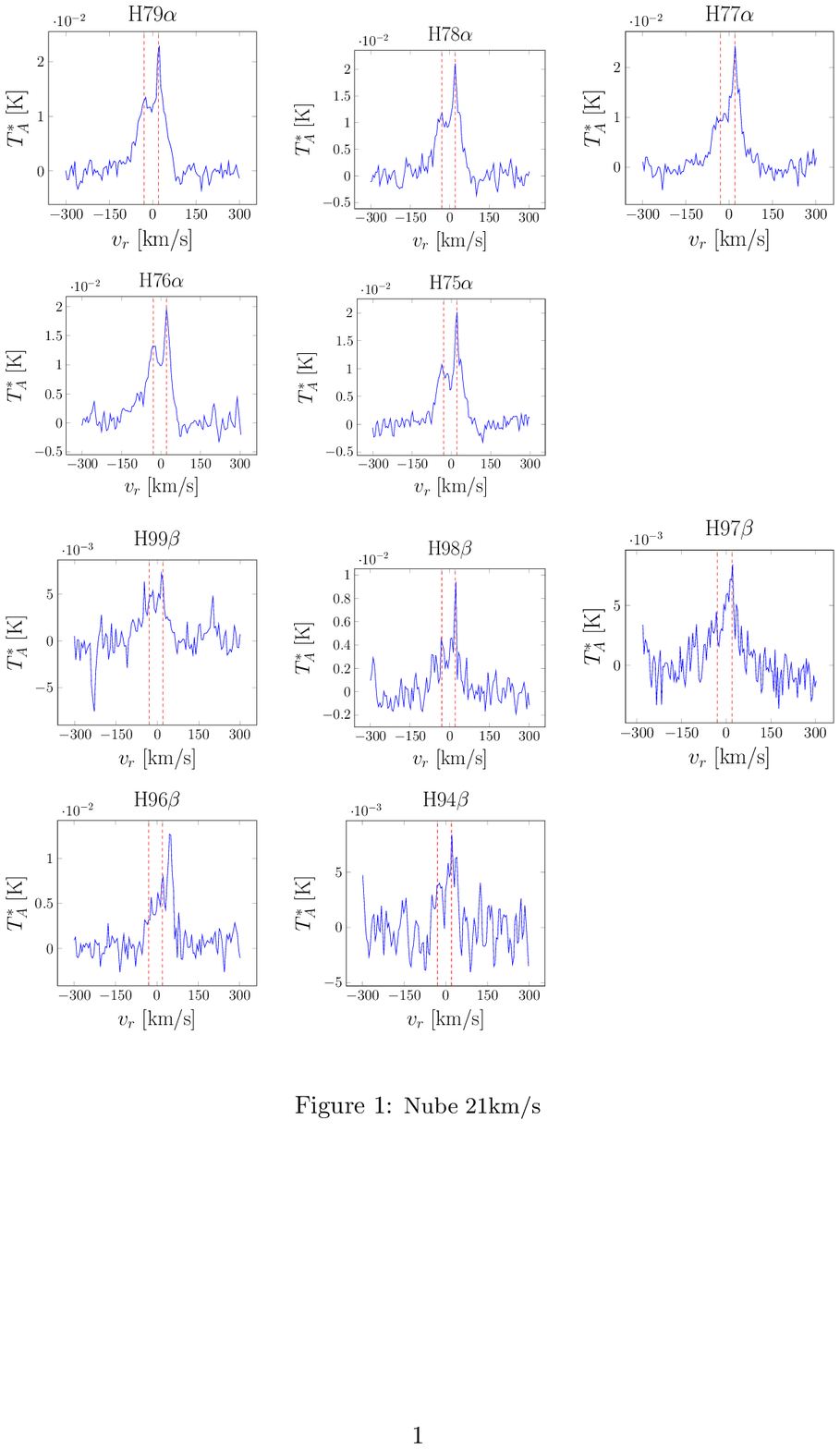}
\caption{Hydrogen RRLs observed toward \emph{LOS}$-$0.11. The dashed red lines show the velocities of 20 and -30 km s$^{-1}$. The spectra are shown with a spectral resolution of \mbox{$\sim$5 km s$^{-1}$}.}\label{fig_RRL_LOS011}
\end{figure}

\begin{figure}
\includegraphics[trim=0 110 0 0, clip, width=\columnwidth]{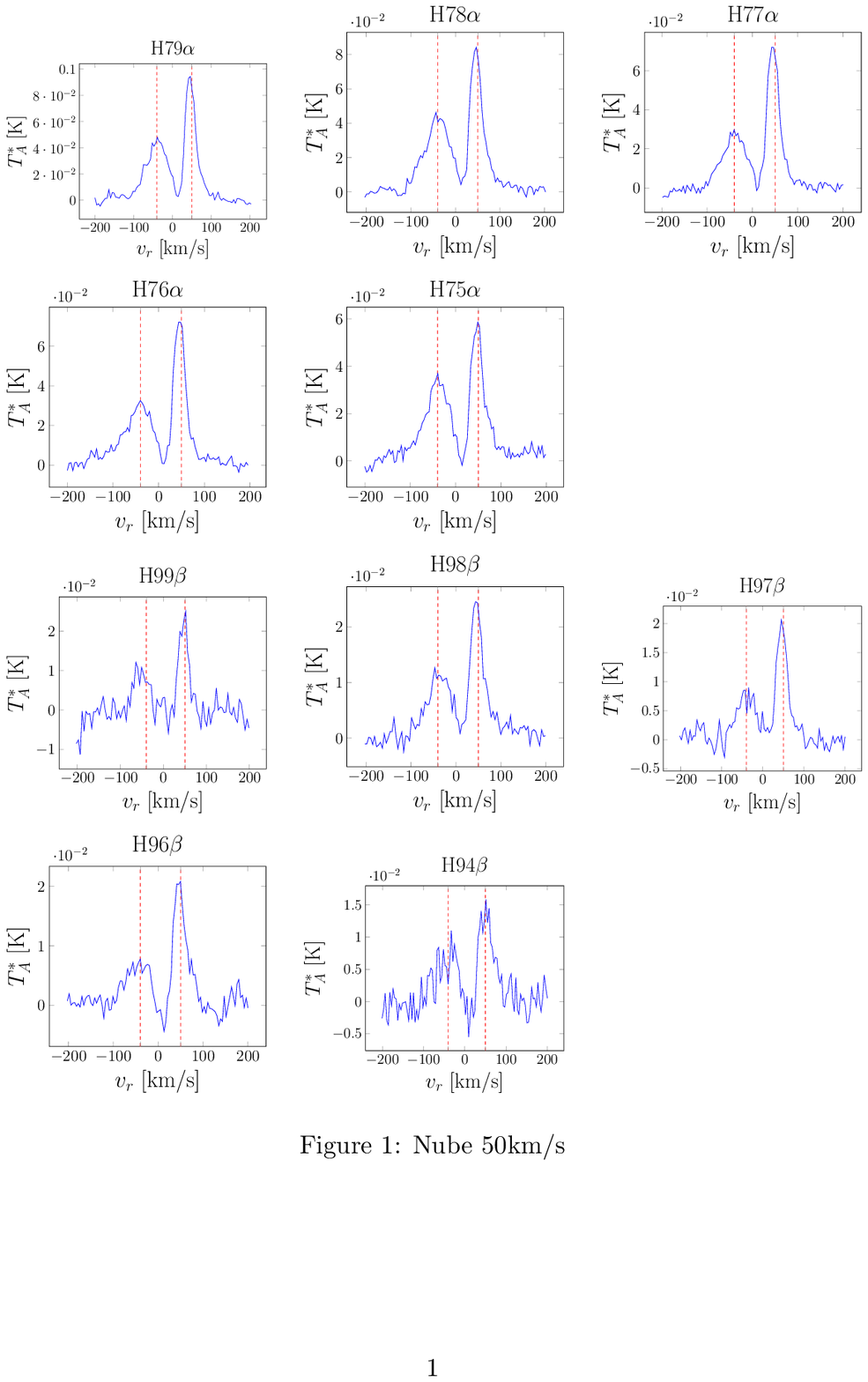}
\caption{Hydrogen RRLs observed toward \emph{LOS}$-$0.02. The dashed red lines show the velocities of 50 and -40 km s$^{-1}$. The spectra are shown with a spectral resolution of \mbox{$\sim$5 km s$^{-1}$}.}\label{fig_RRL_LOS002}
\end{figure}

\begin{figure}
\includegraphics[trim=0 120 0 0, clip, width=\columnwidth]{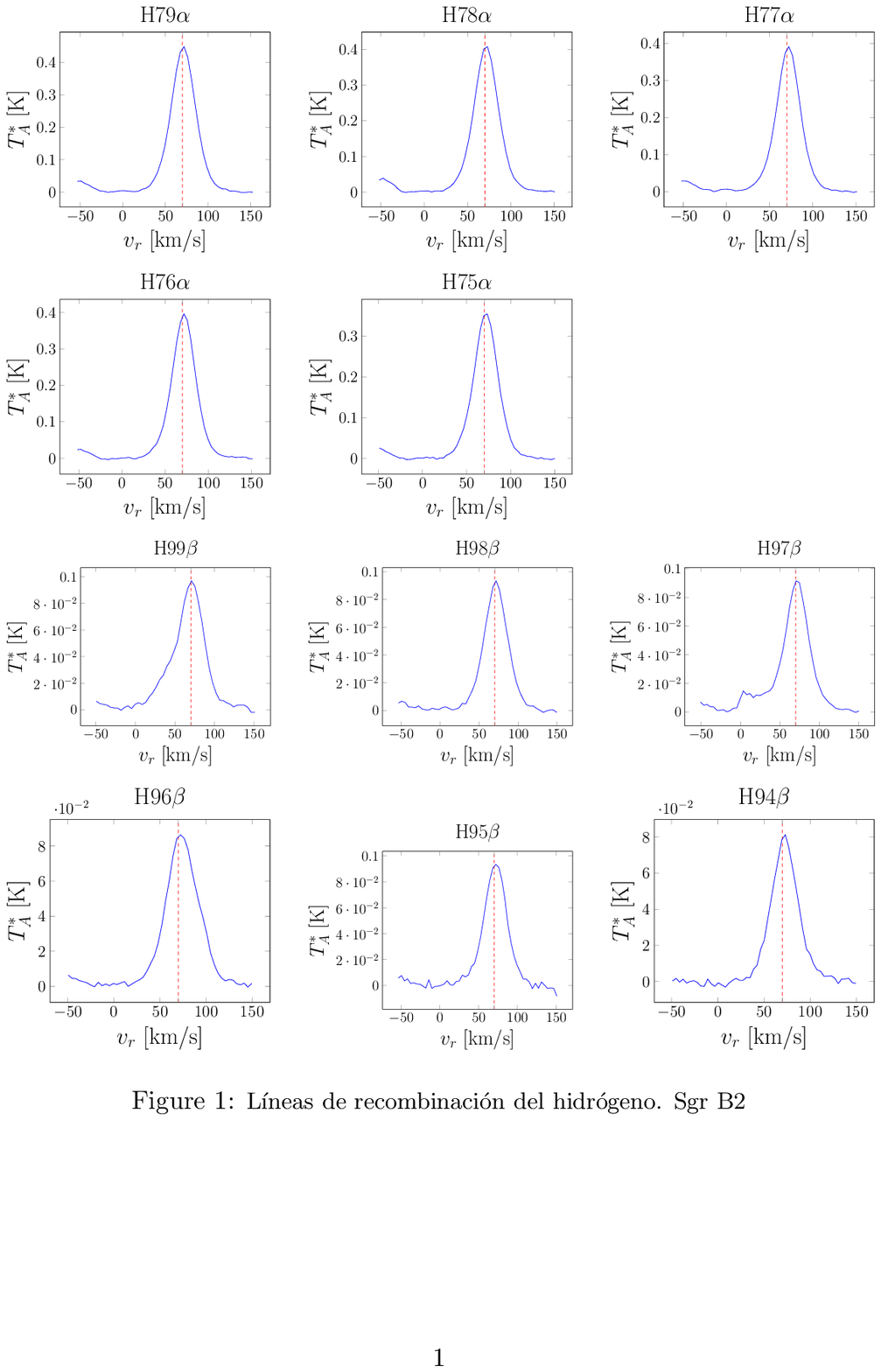}
\caption{Hydrogen RRL observed toward \emph{LOS}+0.693. The dashed red line shows the velocity of 70 km s$^{-1}$. The spectra are shown with a spectral resolution of \mbox{$\sim$5 km s$^{-1}$}.}\label{fig_RRL_LOS0693}
\end{figure}

\begin{figure}
\includegraphics[trim=0 110 0 0, clip, width=\columnwidth]{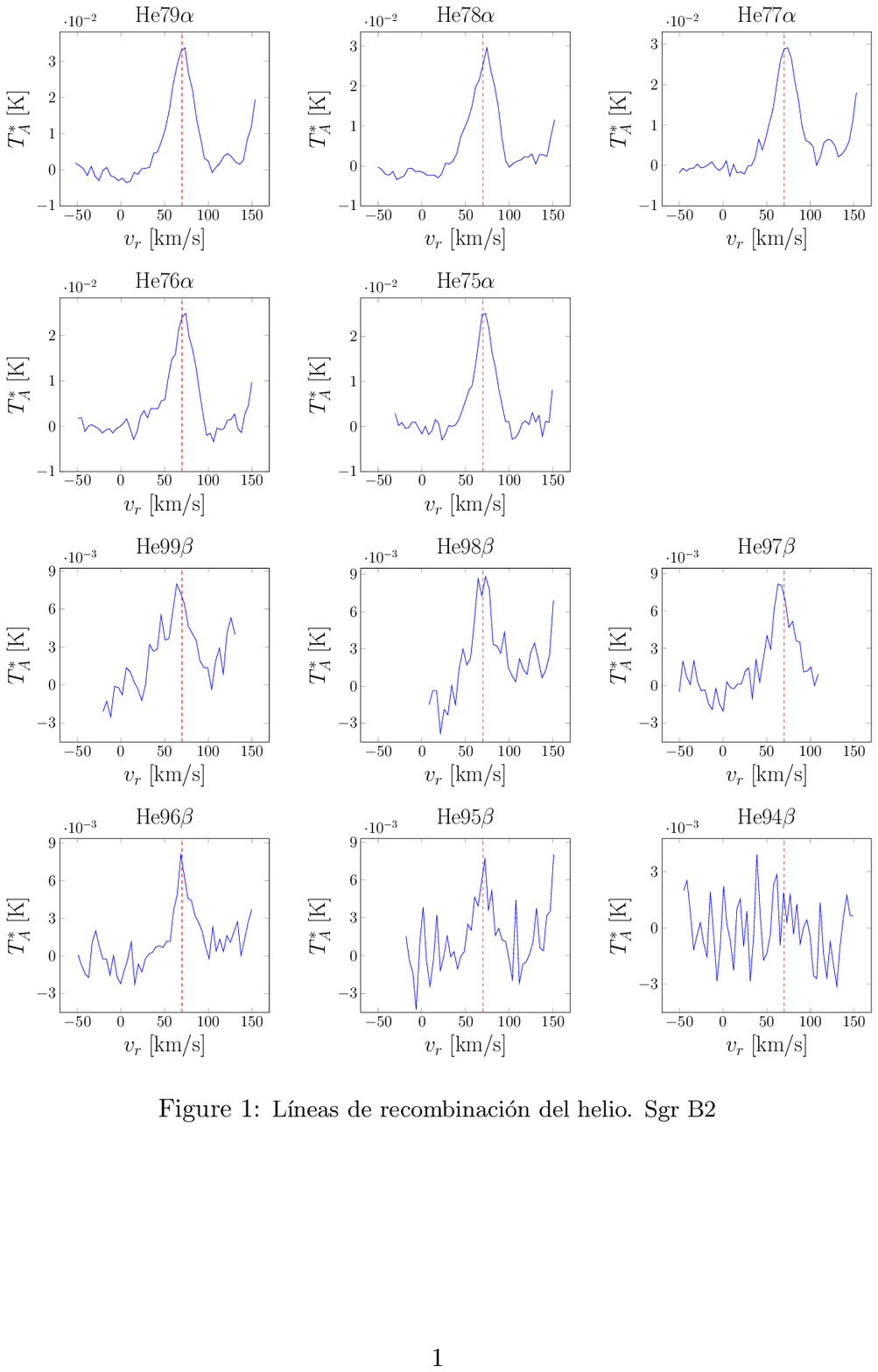}
\caption{Helium RRLs observed toward \emph{LOS}+0.693. The dashed red line shows the velocity of 70 km s$^{-1}$. The spectra are shown with a spectral resolution of \mbox{$\sim$5 km s$^{-1}$}.}\label{fig_RRLHe_LOS0693}
\end{figure}

To identify hydrogen (H) and helium (He) RRLs we have used a catalog included in the MADCUBAIJ package, which contains the frequencies of the RRLs estimated according to the Dirac theory described by \cite{Towle96}. The RRLs detected in \emph{LOS}$-$0.11, \emph{LOS}$-$0.02 and \emph{LOS}+0.693 are shown in Fig.~\ref{fig_RRL_LOS011},~\ref{fig_RRL_LOS002} and~\ref{fig_RRL_LOS0693}-\ref{fig_RRLHe_LOS0693}, respectively. We have detected emission from H$n\alpha$ lines with n=79$-$75 and H$m\beta$ lines with m=99$-$96, 94 toward \emph{LOS}$-$0.11 and \emph{LOS}$-$0.02. Hydrogen RRLs with the same n and m are also detected toward \emph{LOS}+0.693 but, in this case, we have also detected the H95$\beta$, whose frequency is not in the bandwidth of our observations toward either of the two Sgr A sources. All these RRLs are detected with a significance higher than 3$\sigma$. 
The strongest RRLs are observed in \emph{LOS}+0.693, whereas the weakest lines are detected in \emph{LOS}$-$0.11. We have detected the emission from
He$n\alpha$ lines, with n=79$-$75 and He$m\beta$ lines with m=99$-$94, only toward \emph{LOS}+0.693 (see Fig.~\ref{fig_RRLHe_LOS0693}). 
As shown in Fig.~\ref{fig_RRL_LOS011} and~\ref{fig_RRL_LOS002} the RRLs in both of the of \mbox{Sgr A} sources reveal two velocity components, while the RRLs in \emph{LOS}+0.693 show only a single velocity component.

\begin{table}
\scriptsize
\caption{Hydrogen RRL parameters derived for \emph{LOS}$-$0.11 using Gaussian fits with two velocity components.}\label{table1}
\begin{tabular}{ccrrrc}
\hline
  RRL & $\nu$ & \multicolumn{1}{c}{T$^*_\mathrm{A}$} & \multicolumn{1}{c}{$v_\mathrm{r}$} &\multicolumn{1}{c}{$\Delta v_\mathrm{r}$} &\multicolumn{1}{c}{$\int T_\mathrm{A}^* dv_\mathrm{r}$} \\
      & (GHz)  & \multicolumn{1}{c}{(mK)}             &(km s$^{-1}$)        &(km s$^{-1}$)        &($\times$10$^2$ mK km s$^{-1}$)\\
\hline
H79$\alpha$ &13.09 & 19.4$\pm$1.6 &  22.9$\pm$1.3 & 30.2$\pm$2.9 & 6.2$\pm$0.8\\
            &      & 12.6$\pm$1.3 & -26.5$\pm$2.5 &	49.7$\pm$6.5 & 6.7$\pm$1.2\\ 
H78$\alpha$ &13.60 & 17.9$\pm$0.9 &  21.9$\pm$1.3 & 38.2$\pm$2.8 & 7.3$\pm$0.7\\ 
            &      & 11.0$\pm$0.8 & -32.3$\pm$2.4 & 48.6$\pm$5.7 & 5.7$\pm$0.8\\ 
H77$\alpha$ &14.12 & 22.9$\pm$3.5 &  21.8$\pm$0.8 & 27.3$\pm$2.6 & 6.6$\pm$1.3\\
            &      &  9.8$\pm$1.1 & -17.2$\pm$9.1 & 80$\pm$17& 8.4$\pm$2.2\\  
H76$\alpha$ &14.69 & 19.0$\pm$1.1 &  22.9$\pm$1.2 & 29.9$\pm$2.7 & 6.1$\pm$0.7\\
            &      & 13.1$\pm$1.0 & -28.7$\pm$1.9 & 51.3$\pm$4.8 & 7.2$\pm$0.9\\ 
H75$\alpha$ &15.28 & 20.0$\pm$1.0 &  22.5$\pm$0.9 & 30.0$\pm$2.0 & 6.4$\pm$0.6\\
            &      & 10.4$\pm$0.8 & -32.5$\pm$2.0 & 51.6$\pm$5.3 & 5.7$\pm$0.8\\ 
\hline
H99$\beta$  &13.15 & 5.9$\pm$1.1  &  17.3$\pm$3.5 & 25.8$\pm$7.1 & 1.6$\pm$0.6\\
            &      & 4.7$\pm$0.5  & -30.0$\pm$6.0 & 50$\pm$13 & 2.5$\pm$0.8\\ 
H98$\beta$  &13.56 & 9.0$\pm$0.9  &  20.2$\pm$1.7 & 25.1$\pm$4.2 & 2.4$\pm$0.5\\ 
            &      & 3.3$\pm$0.6  & -30.0$\pm$5.2 & 50$\pm$14 & 1.8$\pm$0.6\\ 
H97$\beta$  &13.98 & 8.1$\pm$1.0  &  17.4$\pm$1.6 & 37.0$\pm$3.8 & 3.2$\pm$0.6\\
            &      & 3.0$\pm$0.5  & -30.0$\pm$8.7 & 50$\pm$19 & 1.6$\pm$0.7\\ 
H96$\beta$  &14.41 & 7.6$\pm$0.9  &  16.9$\pm$2.4 & 29.0$\pm$8.3 & 2.3$\pm$0.8\\
            &      & 3.7$\pm$0.6  & -30.0$\pm$7.3 & 50$\pm$16 & 1.9$\pm$0.7\\  
H94$\beta$  &15.34 & 5.7$\pm$0.6  &  22.4$\pm$2.6 & 35.8$\pm$6.1 & 2.1$\pm$0.5\\
            &      & 4.0$\pm$1.0  & -30.0$\pm$3.7 & 50.0$\pm$9.2 & 2.2$\pm$0.8\\ 
\hline 
\end{tabular}
\end{table}

\begin{table}
\scriptsize
\caption{Hydrogen RRL parameters derived for \emph{LOS}$-$0.02 using Gaussian fits with two velocity components.}\label{table2}
\begin{tabular}{ccrrrc}
\hline
  RRL & $\nu$ &\multicolumn{1}{c}{T$^*_\mathrm{A}$} & \multicolumn{1}{c}{$v_\mathrm{r}$} &\multicolumn{1}{c}{$\Delta v_\mathrm{r}$} &\multicolumn{1}{c}{$\int T_\mathrm{A}^* dv_\mathrm{r}$} \\
      &(GHz)  & \multicolumn{1}{c}{(mK)}    &(km s$^{-1}$)        &(km s$^{-1}$)        &($\times$10$^2$ mK km s$^{-1}$)\\
\hline
H79$\alpha$& 13.09 & 92.8$\pm$3.5 &  46.1$\pm$0.5 & 27.6$\pm$1.1 & 27.2$\pm$1.7\\ 
           &       & 45.6$\pm$2.3 & -39.2$\pm$1.4 & 60.6$\pm$3.4 & 29.4$\pm$2.3\\
H78$\alpha$& 13.60 & 83.0$\pm$2.9 &  46.6$\pm$0.5 & 29.7$\pm$1.1 & 26.3$\pm$1.4\\
           &       & 42.9$\pm$2.0 & -41.0$\pm$1.3 & 61.9$\pm$3.1 & 28.2$\pm$2.0\\ 
H77$\alpha$& 14.12 & 72.8$\pm$1.7 &  45.6$\pm$0.4 & 31.8$\pm$0.9 & 24.6$\pm$1.0\\ 
           &       & 28.1$\pm$1.2 & -39.8$\pm$1.3 & 61.8$\pm$3.3 & 18.5$\pm$1.3\\ 
H76$\alpha$& 14.69 & 73.6$\pm$2.1 &  46.5$\pm$0.4 & 29.5$\pm$1.0 & 23.1$\pm$1.1\\ 
           &       & 30.8$\pm$1.5 & -41.6$\pm$1.4 & 63.3$\pm$3.2 & 20.8$\pm$1.5\\ 
H75$\alpha$& 15.28 & 57.4$\pm$2.1 &  48.1$\pm$0.6 & 32.1$\pm$1.3 & 19.6$\pm$1.2\\ 
           &       & 34.6$\pm$1.6 & -39.8$\pm$1.2 & 57.0$\pm$2.9 & 21.0$\pm$1.5\\ 
\hline
H99$\beta$& 13.15  & 23.1$\pm$2.2 &  48.1$\pm$1.2 & 24.4$\pm$2.8 &  6.0$\pm$0.9\\ 
          &        &  9.6$\pm$1.5 & -40.0$\pm$4.1 & 60.0$\pm$9.6 & 6.2$\pm$1.4\\ 
H98$\beta$& 13.56  & 24.4$\pm$1.1 &  45.6$\pm$0.6 & 31.4$\pm$1.5 & 8.2$\pm$0.6\\ 
          &        & 11.1$\pm$0.8	& -40.3$\pm$1.9 & 59.6$\pm$4.4 &  7.1$\pm$0.8\\ 
H97$\beta$& 13.98  & 20.2$\pm$0.9 &  46.0$\pm$0.6 & 28.9$\pm$1.4 &  6.2$\pm$0.4\\
          &        &  7.9$\pm$0.7	& -37.9$\pm$2.0 & 53.6$\pm$4.7 &  4.5$\pm$0.6\\  
H96$\beta$& 14.41  & 20.3$\pm$1.0 &  48.3$\pm$0.7 & 30.5$\pm$1.6 &  6.6$\pm$0.5\\ 
          &        &  8.9$\pm$0.9	& -31.0$\pm$2.8 & 56.2$\pm$7.1 &  5.3$\pm$0.6\\ 
H94$\beta$& 15.34  & 14.6$\pm$1.3 &  51.4$\pm$1.4 & 35.4$\pm$3.4 &  5.5$\pm$0.8\\
          &        &  8.3$\pm$1.0	& -40.2$\pm$2.8 & 60.0$\pm$6.6 &  5.3$\pm$0.9\\  
\hline
\end{tabular}
\end{table}

\begin{table}
\scriptsize
\caption{Hydrogen RRL parameters derived for \emph{LOS}+0.693 using Gaussian fits.}\label{table3}
\begin{tabular}{ccrrrc}
\hline
  RRL & $\nu$ & \multicolumn{1}{c}{T$^*_\mathrm{A}$} & \multicolumn{1}{c}{$v_\mathrm{r}$} &\multicolumn{1}{c}{$\Delta v_\mathrm{r}$} &\multicolumn{1}{c}{$\int T_\mathrm{A}^* dv_\mathrm{r}$} \\
      &(GHz)  &\multicolumn{1}{c}{(mK)}  &(km s$^{-1}$)        &(km s$^{-1}$)        &($\times$10$^3$ mK km s$^{-1}$)\\
\hline  
H79$\alpha$& 13.09  & 441.6$\pm$3.2 & 71.4$\pm$0.1  & 31.0$\pm$0.3 & 14.6$\pm$0.2\\ 
H78$\alpha$& 13.60  & 401.4$\pm$3.1 & 71.8$\pm$0.1  & 32.0$\pm$0.3 & 13.7$\pm$0.2\\ 
H77$\alpha$& 14.12  & 382.0$\pm$4.8 & 71.8$\pm$0.2  & 31.0$\pm$0.4 & 12.6$\pm$0.2\\ 
H76$\alpha$& 14.69  & 383.3$\pm$6.5 & 71.7$\pm$0.2  & 30.7$\pm$0.6 & 12.5$\pm$0.3\\ 
H75$\alpha$& 15.28  & 350.4$\pm$2.5 & 72.3$\pm$0.1  & 30.0$\pm$0.3 & 11.2$\pm$0.1\\ 
\hline 
H99$\beta$&  13.15  & 93.2$\pm$3.1  & 70.2$\pm$0.5  & 37.0$\pm$1.3 & 3.7$\pm$0.2\\ 
H98$\beta$&  13.56  & 91.1$\pm$1.0  & 71.8$\pm$0.3  & 33.0$\pm$0.4 & 3.2$\pm$0.1\\ 
H97$\beta$&  13.98  & 88.4$\pm$3.7  & 71.3$\pm$0.4  & 32.2$\pm$1.0 & 3.0$\pm$0.1\\ 
H96$\beta$&  14.41  & 84.5$\pm$1.8  & 73.5$\pm$0.4  & 37.4$\pm$0.8 & 3.4$\pm$0.1\\ 
H95$\beta$&  14.87  & 91.9$\pm$2.8  & 72.0$\pm$0.4  & 30.5$\pm$1.0 & 3.0$\pm$0.1\\ 
H94$\beta$&  15.34  & 78.5$\pm$1.0  & 72.6$\pm$0.2  & 32.0$\pm$0.5 & 2.7$\pm$0.1\\ 
\hline
\end{tabular}
\end{table}

\begin{table}
\scriptsize
\caption{Helium RRL parameters derived for \emph{LOS}+0.693 using Gaussian fits.}\label{table4}
\begin{tabular}{ccrrrc}
\hline
  RRL & $\nu$ &\multicolumn{1}{c}{T$^*_\mathrm{A}$} & \multicolumn{1}{c}{$v_\mathrm{r}$} &\multicolumn{1}{c}{$\Delta v_\mathrm{r}$} &\multicolumn{1}{c}{$\int T_\mathrm{A}^* dv_\mathrm{r}$} \\
      &(GHz)  &\multicolumn{1}{c}{(mK)}  &(km s$^{-1}$)        &(km s$^{-1}$)        &($\times$10$^2$ mK km s$^{-1}$)\\
\hline 
He79$\alpha$& 13.09   & 31.8$\pm$2.0  & 71.5$\pm$0.7 & 26.6$\pm$1.6  & 9.0$\pm$0.8\\ 
He78$\alpha$& 13.60   & 35.8$\pm$2.9  & 73.5$\pm$0.9 & 26.0$\pm$2.0  & 9.9$\pm$1.2\\ 
He77$\alpha$& 14.13   & 28.2$\pm$3.9  & 73.0$\pm$1.6 & 26.8$\pm$3.7  & 8.0$\pm$1.7\\ 
He76$\alpha$& 14.70   & 23.4$\pm$1.8  & 72.8$\pm$0.8 & 25.2$\pm$1.9  & 6.3$\pm$0.7\\ 
He75$\alpha$& 15.29   & 24.3$\pm$5.4  & 72.3$\pm$2.3 & 23.3$\pm$5.4 & 6.0$\pm$2.0\\ 
\hline 
He99$\beta$&  13.15   &  7.1$\pm$3.5  & 65.7$\pm$5.6 & 30$\pm$13 & 2.3$\pm$1.6\\ 
He98$\beta$&  13.56   &  9.5$\pm$2.1  & 70.8$\pm$1.9 & 22.2$\pm$4.6  & 2.2$\pm$0.7\\ 
He97$\beta$&  13.98   &  8.0$\pm$2.7  & 66.6$\pm$4.5 & 27$\pm$11 & 2.3$\pm$1.3\\ 
He96$\beta$&  14.42   &  7.7$\pm$1.2  & 70.2$\pm$1.3 & 21.7$\pm$3.1  & 1.8$\pm$0.4\\ 
He95$\beta$&  14.87   &  7.8$\pm$1.3  & 71.8$\pm$1.1 & 17.4$\pm$2.5 & 1.4$\pm$0.3\\ 
He94$\beta$&  15.35   & $<$6.5$^{(a)}$ & ... & ... & $<$1.7$^{(b)}$\\
\hline
\end{tabular}
$^{(a)}$3$\sigma$ upper limit on the line intensity.\\
$^{(b)}$3$\sigma$ upper limit on the velocity$-$integrated line intensity.
\end{table}

\begin{table}
\centering
\caption{3$\sigma$ upper limits on the He line intensities for \emph{LOS}$-$0.11 and \emph{LOS}$-$0.02.}\label{table5}
\begin{tabular}{ccrr}
\hline
    &       & \emph{LOS}$-$0.11 & \emph{LOS}$-$0.02\\ 
RRL & $\nu$ & \multicolumn{2}{c}{T$^*_\mathrm{A}$}\\
    & (GHz) & \multicolumn{2}{c}{(mK)}\\
\hline
He79$\alpha$& 13.09 & $<$5  & $<$11\\ 
He78$\alpha$& 13.60 & $<$3  & $<$9\\ 
He77$\alpha$& 14.13 & $<$11 & $<$5\\ 
He76$\alpha$& 14.70 & $<$3  & $<$6\\ 
He75$\alpha$& 15.29 & $<$3  & $<$6\\ 
\hline
He99$\beta$& 13.15  & $<$3  & $<$8\\ 
He98$\beta$& 13.56  & $<$3  & $<$3\\ 
He97$\beta$& 13.98  & $<$3  & $<$3\\ 
He96$\beta$& 14.42  & $<$3  & $<$3\\ 
He94$\beta$& 15.35  & $<$3  & $<$4\\ 
\hline
\end{tabular}
\end{table}

Gaussian fits to the RRLs are used to derive the peak intensity (T$_\mathrm{A}^*$), central line velocity ($v_r$), full width at half maximum ($\Delta$$v_r$), the integrated line intensity ($\int T_\mathrm{A}^* dv_\mathrm{r}$), and their respective uncertainties. The frequencies of the RRLs and the derived parameters for each source are listed in Tables \ref{table1}$-$\ref{table3}. 
The RRLs found in both sources of Sgr A are fitted with two Gaussian lines. The two velocity components are labelled as +20 and -30 km s$^{-1}$ in \emph{LOS}$-$0.11 and as +50 and -40 km s$^{-1}$ in \emph{LOS}$-$0.02 in Tables \ref{table6}, \ref{table10} and \ref{table9}.
As mentioned, He$n\alpha$ lines are detected only in \emph{LOS}+0.693, and the parameters derived using Gaussian fits are given in Table \ref{table4}, where upper
limits for the T$_\mathrm{A}^*$ and  $\int T_\mathrm{A}^* dv_\mathrm{r}$ of the He94$\beta$ line are also listed. For both sources of \mbox{Sgr A} we have estimated 3$\sigma$ upper limits for T$_\mathrm{A}^*$ of the 
He lines shown in Table \ref{table5} because we will study the He to H ratio in \mbox{Section \ref{He_to_H_ratio}.}
In \mbox{Table \ref{table5}} there are no upper limits for the He95$\beta$ line as it was not observed in either of the Sgr A sources.

\subsection{LTE conditions}\label{LTE_conditions}

In order to check whether local thermodynamic equilibrium (LTE) conditions apply in the three GC sources, we have derived the H$m\beta$ to H$n\alpha$ integrated line intensity ratios (hereafter H$m\beta$ to H$n\alpha$ ratios), using H$n\alpha$ and H$m\beta$ lines that were observed simultaneously to avoid uncertainties related to pointing and flux calibration. We show the value of these ratios in \mbox{Table \ref{table6}} for the different velocity components and the three GC sources. In this table we also list the H$m\beta$ to H$n\alpha$ ratios estimated assuming LTE conditions and optically thin radio continuum emission. We also show the He$m\beta$ to He$n\alpha$ integrated line intensity ratios for \emph{LOS}+0.693 in Table \ref{table7}, where the expected LTE values are also listed.

As seen in Table \ref{table6}, the three GC sources have H$m\beta$ to H$n\alpha$ ratios that are consistent, within their uncertainties, with those predicted in LTE, but there are values (in bold print) in this table that do not match the expected LTE ratio. 
In \emph{LOS}$-$0.02, the measured H99$\beta$ to H79$\alpha$ and H98$\beta$ to H78$\alpha$ (positive velocity component) ratios are inconsistent
with the LTE values likely due to uncertainty in the baseline correction of the H99$\beta$ and H98$\beta$ lines. 
The same reason may explain why the H99$\beta$, H98$\beta$, and H94$\beta$ lines, in \emph{LOS}+0.693, show intensities lower than those expected in LTE.
On the other hand, the measured He$m\beta$ to He$n\alpha$ ratios (see Table \ref{table7}) are consistent within their uncertainties with the values expected in LTE. In summary, the H$m\beta$ to H$n\alpha$ ratios derived for the three GC sources and the He$m\beta$ to He$n\alpha$ ratios derived for \emph{LOS}+0.693 show that the ionised gas in the studied sources can be reasonably assumed to be emitted under LTE conditions.

\begin{table*}
\caption{H$m\beta$/H$n\alpha$ ratios for the three GC \emph{LOSs}}\label{table6}
\begin{threeparttable}
\begin{tabular}{cccrcrcr}
\hline
                      && \multicolumn{2}{c}{\emph{LOS}$-$0.11} & \multicolumn{2}{c}{\emph{LOS}$-$0.02} & \multicolumn{2}{c}{\emph{LOS}+0.693}\\
          Ratio       & Model$^\mathrm{(a)}$   & $v_\mathrm{r}^{\mathrm{(b)}}$     & \multicolumn{1}{c}{$\frac{I_{\mathrm{H}m\beta}}{I_{\mathrm{H}n\alpha}}$} & $v_\mathrm{r}^\mathrm{(b)}$     & \multicolumn{1}{c}{$\frac{I_{\mathrm{H}m\beta}}{I_{\mathrm{H}n\alpha}}$$^\mathrm{(c)}$} & $v_\mathrm{r}^{\mathrm{(b)}}$ & \multicolumn{1}{c}{$\frac{I_{\mathrm{H}m\beta}}{I_{\mathrm{H}n\alpha}}$$^\mathrm{(c)}$}\\
                      &  (\%)     & (km s$^{-1}$) &  \multicolumn{1}{c}{(\%)} & (km s$^{-1}$) & \multicolumn{1}{c}{(\%)} & (km s$^{-1}$) & \multicolumn{1}{c}{(\%)} \\      
\hline
H99$\beta$/H79$\alpha$& 27.3& +20 & 25.8$\pm$9.8 &  +50  & \textbf{22.1$\pm$3.7}& +70 & \textbf{25.2$\pm$1.3}\\
                      &     & -30 & 37$\pm$13 &  -40 & \textbf{20.9$\pm$5.2} &  & \\
H98$\beta$/H78$\alpha$& 27.0& +20 & 33.0$\pm$7.6 &  +50  & \textbf{31.1$\pm$2.7} & +70 & \textbf{23.4$\pm$0.5}\\
                      &     & -30 & 31$\pm$11 &  -40 & 25.0$\pm$3.2& & \\
H96$\beta$/H77$\alpha$& 27.7& +20 & 35$\pm$13 &  +50 & 26.8$\pm$2.3& +70 & 26.7$\pm$1.0\\
                      &     & -30 & 23$\pm$11 &  -40 & 28.8$\pm$3.9&    & \\
H96$\beta$/H76$\alpha$& 26.6& +20 & 39$\pm$13 &  +50 & 28.6$\pm$2.5& +70 & 26.8$\pm$1.1\\
                      &     & -30 & 27$\pm$11 &  -40 & 25.7$\pm$3.5&    & \\
H94$\beta$/H75$\alpha$& 26.6& +20 & 33.3$\pm$7.7 &  +50  & 28.2$\pm$4.2& +70 & \textbf{23.9$\pm$0.6}\\
                      &     & -30 & 38$\pm$14 &  -40 & 25.2$\pm$4.6&    & \\
\hline
\end{tabular}
\begin{tablenotes}
\item $^{(a)}$ Estimated values assuming LTE conditions and optically thin radio continuum emission (see text).
\item $^{(b)}$ The velocity components identified in \emph{LOS}$-$0.11 and \emph{LOS}$-$0.02 are labelled as +20 and -30 km s$^{-1}$, and +50 and -40 km s$^{-1}$, respectively. Only one velocity component labelled as +70 km s$^{-1}$ is identified in \emph{LOS}+0.693.
\item $^{(c)}$ Values or limits that do not match the expected LTE ratio are in bold print.  
\end{tablenotes}
\end{threeparttable}
\end{table*}

\begin{table}
\centering
\caption{He$m\beta$/He$n\alpha$ ratios for \emph{LOS}+0.693}\label{table7}
\begin{threeparttable}
\begin{tabular}{ccr}
\hline
 Ratio & Model$^{(a)}$ & \multicolumn{1}{c}{$\frac{I_{\mathrm{He}m\beta}}{I_{\mathrm{He}n\alpha}}$}     \\
 & (\%) & \multicolumn{1}{c}{(\%)} \\
\hline
He99$\beta$/He79$\alpha$& 27.3  &25$\pm$18 \\ 
He98$\beta$/He78$\alpha$& 27.0  &22.7$\pm$7.8 \\ 
He96$\beta$/He77$\alpha$& 27.7  &22.1$\pm$6.7 \\ 
He96$\beta$/He76$\alpha$& 26.6  &28.3$\pm$7.0 \\ 
He94$\beta$/He75$\alpha$& 26.6  &$<$25.4\\
\hline
\end{tabular}
\begin{tablenotes}
\item $^{(a)}$ Estimated values assuming LTE conditions and optically thin radio continuum emission (see text).
\end{tablenotes}
\end{threeparttable}
\end{table}


\subsection{Helium to hydrogen ratio}\label{He_to_H_ratio}

As mentioned above, helium RRLs have only been detected toward \emph{LOS}+0.693. We have derived the He$-$to$-$H line intensity ratios (see \mbox{Table \ref{table10})} for those RRL transitions where the same principal quantum number has been detected for the two elements. Otherwise, we provide upper limits assuming that the line intensity of the non$-$detected RRLs is lower than 3$\sigma$. We note that the He$n\alpha$ RRLs are located at \mbox{$\approx$122 km s$^{-1}$} with respect to H$n\alpha$ RRLs, as expected by the difference of their rest frequencies \citep{Towle96}. Thus, the derived ratios are not affected by possible effects of calibration since both spectral lines are observed simultaneously at close frequencies.

We find an average He$-$to$-$H intensity ratio of 7.3$\pm$0.2 per cent or $^4$He mass fraction Y=0.29$\pm$0.01 for \emph{LOS}+0.693. This ratio is consistent with those found in interferometry studies \citep{Roelfsema86,Mehringer93} that trace more compact regions ($\lesssim$0.7 \textrm{pc}) than our diffuse regions ($\sim$1.7 \textrm{pc}). The most stringent upper limits on the He$-$to$-$H ratio derived for \emph{LOS}$-$0.02 and
\emph{LOS}$-$0.11 are consistent with the He$-$to$-$H number ratio of $<$10 per cent found in GC H{\small \,II} regions \citep{Roelfsema86}. The estimation of 0.29$\pm$0.01 helium abundance by mass differs by 14 per cent from that of 0.25 as predicted by Big Bang nucleosynthesis \citep{Coc12,Tsivilev13}. This finding suggests, as expected \citep{Wilson94,Gordon09}, that high$-$mass stars in the 
GC have enriched the ISM with helium$-$4, in a past intense burst of star formation in this region, thus increasing its abundance compared to the primordial value.

\begin{table*}
\centering
\caption{Helium$-$to$-$hydrogen line intensity ratios for the three GC sources}\label{table10}
\begin{threeparttable}
\begin{tabular}{cccccccr}
\hline
                        & \multicolumn{2}{c}{\emph{LOS}$-$0.11} & \multicolumn{2}{c}{\emph{LOS}$-$0.02} & \multicolumn{2}{c}{\emph{LOS}+0.693}\\
  Ratio                 & v$_\mathrm{r}^{(a)}$  & $\frac{I_{\mathrm{He}n\alpha}}{I_{\mathrm{H}n\alpha}}$ & v$_\mathrm{r}^{(a)}$ & $\frac{I_{\mathrm{He}n\alpha}}{I_{\mathrm{H}n\alpha}}$ & v$_\mathrm{r}^{(a)}$ & \multicolumn{1}{c}{$\frac{I_{\mathrm{He}n\alpha}}{I_{\mathrm{H}n\alpha}}$} \\
                        & (km s$^{-1}$)&   (\%)        & (km s$^{-1}$) &   (\%)      & (km s$^{-1}$) & (\%) \\
\hline
He79$\alpha$/H79$\alpha$& +20  & $<$20 & +50 & $<$10   & +70&7.2$\pm$0.5 \\
                        & -30  & $<$40 & -40 & $<$20   & &  \\
He78$\alpha$/H78$\alpha$& +20  & $<$10 & +50 & $<$10   & +70&9.0$\pm$0.7 \\ 
                        & -30  & $<$20 & -40 & $<$20   & &  \\
He77$\alpha$/H77$\alpha$& +20  & $<$50 & +50 & $<$10   & +70&7.4$\pm$1.0 \\  
                        & -30  & $<$100 & -40 &$<$20   & &  \\
He76$\alpha$/H76$\alpha$& +20  & $<$20 & +50 & $<$10   & +70&6.1$\pm$0.5 \\ 
                        & -30  & $<$30 & -40 & $<$20   & &  \\
He75$\alpha$/H75$\alpha$& +20  & $<$20 & +50 & $<$10   & +70&6.9$\pm$1.5 \\ 
                        & -30  & $<$30 & -40 & $<$20   & &  \\
\hline 
He99$\beta$/H99$\beta$  & +20  &$<$50 & +50 &$<$40 & +70&7.6$\pm$3.8 \\
                        & -30  &$<$70 & -40 &$<$90 & &\\ 
He98$\beta$/H98$\beta$  & +20  &$<$30 & +50 &$<$10 & +70&10.5$\pm$2.3 \\
                        & -30  &$<$80 & -40 &$<$30 & &\\ 
He97$\beta$/H97$\beta$  & +20  &$<$40 & +50 &$<$10 & +70&9.0$\pm$3.1 \\
                        & -30  &$<$100& -40 &$<$30 & &\\
He96$\beta$/H96$\beta$  & +20  &$<$40 & +50 &$<$10 & +70&9.1$\pm$1.4 \\
                        & -30  &$<$80 & -40 &$<$30 & &\\
He95$\beta$/H95$\beta$$^{(b)}$ & +20 & ...     & +50 & ... & +70 &8.5$\pm$1.4 \\
                        & -30  & ...           & -40 & ...   & &\\
He94$\beta$/H94$\beta$  & +20  &$<$60 & +50 &$<$30 & +70&$<$8.3\\
                        & -30  &$<$90 & -40 &$<$50 & &\\
\hline
\end{tabular}
\begin{tablenotes}
\item $^{(a)}$The velocity components identified in \emph{LOS}$-$0.11 and \emph{LOS}$-0.02$ are labelled as +20 and -30 km s$^{-1}$, and +50 and -40 km s$^{-1}$, respectively. Only one velocity component labelled as +70 km s$^{-1}$ is identified in \emph{LOS}+0.693.
\item $^{(b)}$The He95$\beta$/H95$\beta$ line intensity ratios for both sources of Sgr A were not measured because the He95$\beta$ RRL was not observed.
\end{tablenotes}
\end{threeparttable}
\end{table*}

\subsection{Electron densities and number of Lyman ionising photons}\label{Physical_prop}

In this section we derive the average electron density $n_\mathrm{e}$ of the ionised gas following the equation as in \cite{Mezger67}, where $n_\mathrm{e}$ is given by:
\begin{equation}\label{equa1}
\begin{array}{rl}
\left(\dfrac{n_\mathrm{e}}{\mathrm{cm^{-3}}}\right) =&6.351 \cdot 10^2 u_1 a^{0.5}  \left(\dfrac{T_\mathrm{e}}{10^4 \mathrm{\ K}}\right)^{0.175} \left(\dfrac{\nu}{\mathrm{GHz}}\right)^{0.05}  \\
&\left(\dfrac{S_\mathrm{c}}{\mathrm{Jy}}\right)^{0.5} \left(\dfrac{D}{\mathrm{kpc}}\right)^{-0.5} \left(\dfrac{\Theta}{\mathrm{arcmin}}\right)^{-1.5} \mathrm{,}
\end{array}
\end{equation}

\noindent where $T_\mathrm{e}$ is the electron temperature, $\nu$ is the frequency, $S_{\mathrm{c}}$ is the continuum flux, $D$ is the distance to the GC (\mbox{7.86 \textrm{kpc}}, \cite{Boehle16}) and $\Theta$ is the source size (which is assumed to be equal to the telescope beam size corresponding to \mbox{$\approx$1.7 pc} at the GC distance). The parameter $a$ accounts for the deviation between the exact equation for the optical depth for free$-$free emission and its approximation \citep{Mezger67}. For our study we have used an average value of $a$ equal to 0.98. We have assumed that our three GC sources have spherical geometry, and in this case the model conversion factor u$_1$ is equal to 0.775 \citep{Mezger67}. The $S_{\mathrm{c}}$ is derived from the H77$\alpha$ and H96$\beta$ RRL emission assuming an optically thin regime and the average $T_\mathrm{e}$ found for compact H{\small \,II} regions of the GC, i.e. $T_\mathrm{e} \approx 6300$ K \citep{Goss85}. $S_{\mathrm{c}}$ values derived from the H77$\alpha$ and H96$\beta$ RRLs are similar, within their uncertainties, to those obtained from the other detected Hn$\alpha$ and Hm$\beta$ lines, respectively. For this reason Table \ref{table9} lists only the $S_{\mathrm{c}}$ values derived from the H77$\alpha$ and H96$\beta$ lines.
For the three GC sources the estimated values of n$_\mathrm{e}$ are given in \mbox{Table \ref{table9}.}

Using the formula given in \cite{Rohlfs} (see \mbox{Eq. 13.2)} we have also calculated the number of Lyman ionising photons, N$_\mathrm{Lyc}$, as follows:

\begin{equation}\label{equa2}
N_\mathrm{Lyc}=\dfrac{4}{3}\pi\left(\dfrac{\Theta}{2}\right)^3 n_\mathrm{e} n_\mathrm{p} \alpha^{(2)} \mathrm{,}
\end{equation}

\noindent where $n_\mathrm{p}$ is the proton density (which is 
equal to $n_\mathrm{e}$ under LTE conditions, see \mbox{Section \ref{LTE_conditions}),} and $\alpha^{(2)}$ is the recombination coefficient \citep{Spitzer04}.
The derived values of $N_\mathrm{Lyc}$ for the three GC sources are shown in Table \ref{table9}.

\begin{table}
\tiny
\caption{Physical properties derived for the three GC sources}\label{table9}
\begin{threeparttable}
\begin{tabular}{cccrrr}
\hline
Source & RRL & v$_\mathrm{r}^{(a)}$ & \multicolumn{1}{c}{S$_\mathrm{c}$} & \multicolumn{1}{c}{n$_\mathrm{e}$} & \multicolumn{1}{c}{$\log(N_\mathrm{Lyc})$} \\
& & (km s$^{-1}$) & \multicolumn{1}{c}{(mJy)} & \multicolumn{1}{c}{(cm$^{-3}$)} & \multicolumn{1}{c}{(ph. s$^{-1}$)} \\
\hline
\emph{LOS}$-$0.11 & H77$\alpha$ & +20 & 122$\pm$23   & 71$\pm$7 & 47.14$\pm$0.08\\
&  & -30 & 153$\pm$40 & 80$\pm$10 & 47.24$\pm$0.10 \\
& H96$\beta$  & +20 & 42$\pm$14  & 43$\pm$7 & 46.70$\pm$0.12 \\
&  & -30 & 35$\pm$13 & 39$\pm$7 & 46.62$\pm$0.14 \\
\hline
\emph{LOS}$-$0.02 & H77$\alpha$ & +50 & 451$\pm$18 & 137$\pm$3 & 47.71$\pm$0.02\\
&  & -40 & 339$\pm$25 & 119$\pm$4 & 47.58$\pm$0.03 \\
& H96$\beta$ & +50 & 118$\pm$9  & 72$\pm$3 & 47.15$\pm$0.03 \\
&  & -40 & 96$\pm$11 & 65$\pm$4 & 47.06$\pm$0.05\\
\hline
\emph{LOS}+0.693 & H77$\alpha$ & +70 & 2311$\pm$45 & 310$\pm$3 & 48.42$\pm$0.01 \\
& H96$\beta$  & +70 & 603$\pm$19 &163$\pm$3 & 47.86$\pm$0.01\\
\hline
\end{tabular}
\begin{tablenotes}
\item $^{(a)}$The velocity components identified in \emph{LOS}$-$0.11 and \emph{LOS}$-0.02$ are labelled as +20 and -30 km s$^{-1}$, and +50 and -40 km s$^{-1}$, respectively. Only one velocity component labelled as +70 km s$^{-1}$ is identified in \emph{LOS}+0.693.
\end{tablenotes}
\end{threeparttable}
\end{table}

\section{Discussion}

\subsection{Extended and diffuse RRL emission toward the three GC \emph{LOS}}\label{Extended_emission}

\begin{figure}
\includegraphics[width=\columnwidth]{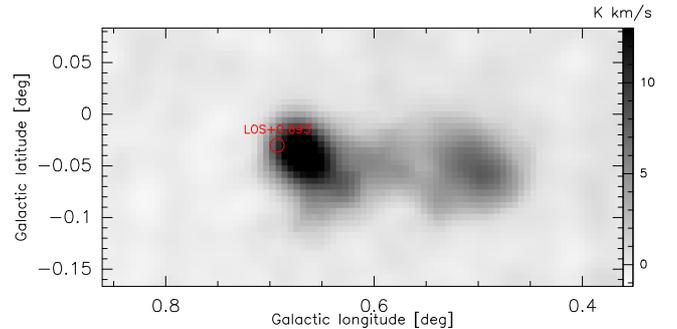}
\caption{H69$\alpha$ integrated line emission of Sgr B2 obtained using HOPS data \citep{Purcell12}. The range of velocity integration is from 20 to 80 km s$^{-1}$. \emph{LOS}+0.693 is indicated with a red circle with the size of the HPBW of the GBT observations (48 arcsec at 13.09 GHz).}\label{SgrB2_H69a}
\end{figure}

As previously mentioned, the only compact H{\small \,II} region which falls inside the GBT beam of our observations is toward \emph{LOS}$-$0.02. This suggests that our GBT observations trace extended RRL emission toward \emph{LOS}$-$0.11 and \emph{LOS}+0.693. In the case of \emph{LOS}+0.693, this idea is also supported by the extended H69$\alpha$ emission map of Sgr B2 shown in \mbox{Fig.~\ref{SgrB2_H69a}.} This figure is obtained using the HOPS data \citep{Purcell12}. The HOPS data has a spatial resolution of 2.4 arcmin at the frequency (19.59 GHz) of the H69$\alpha$ line, which is a factor $\sim$3 worse than the average spatial resolution of our observations. 
Unfortunately, the HOPS data has a rms noise of $\sim$40 mK, which is not enough to obtain H69$\alpha$ line emission maps for regions where both Sgr A sources were observed.

\begin{figure}
\includegraphics[width=\columnwidth]{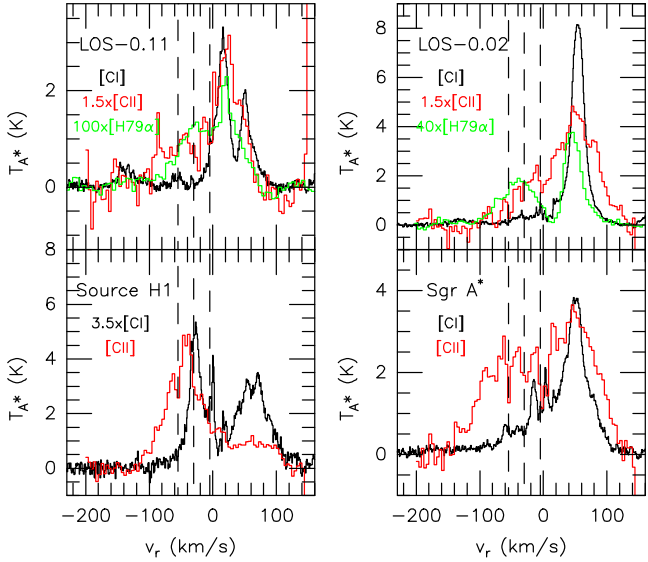}
\caption{\textbf{Upper panels:} C{\small \,I} (black), C{\small \,II} (red) and H79$\alpha$ (green) spectra observed toward \emph{LOS}$-$0.11 and \emph{LOS}$-$0.02. \mbox{\textbf{Bottom panels:}} C{\small \,I} (black) and C{\small \,II} (red) spectra observed toward the H1 source and Sgr A*. The line intensity of several spectra is multiplied by a factor for comparison purposes. C{\small \,II} spectra is affected by absorption features (indicated by dashed lines) associated with the 3 kpc, 4.5 kpc and local spiral arms \citep{Oka98}. The C{\small \,I} and C{\small \,II} spectra are obtained with a spatial resolution of 46 arcsec similar to that of the H79$\alpha$ spectra.}\label{CII_RRL_spectra}
\end{figure}

In order to figure out whether the emission detected by the GBT toward \emph{LOS}$-$0.02 is arising exclusively from the compact A region or not, we have compared the RRL emission measured using the VLA with that of our GBT observations. For this we have first determined the spectral index $\alpha$ of the region A. This region shows a $\alpha$ of 0.06$\pm$0.04 derived considering the $S_{\mathrm{c}}$ of 590$\pm$30 mJy that we have measured at 24.5 GHz (using the VLA map shown in Fig.~\ref{fig1}) and the value of 570$\pm$20 mJy derived at 14.7 GHz by \cite{Goss85}. Thus, T$^*_\mathrm{A}$ $\propto$ $\nu^{1.16}$ assuming that the free$-$free and RRL emission is optically thin. We have measured the H64$\alpha$ peak line intensity of 67$\pm$22 mJy by integrating the channel map at the peak intensity of 47 km s$^{-1}$ (obtained using the data cube described in Section \ref{subsection_of_archival_VLA_data}) over the HPBW of the GBT. By using the previous relation we have extrapolated the H64$\alpha$ peak line intensity to that expected for the H76$\alpha$ line, finding a T$^*_\mathrm{A}$ of 39$\pm$12 mJy at 14.7 GHz, which is similar to that of 40.2$\pm$1.2 mJy as measured by the GBT. This suggests that part of the region A inside the GBT beam may contribute significantly to the RRL emission detected in \emph{LOS}$-$0.02. The T$^*_\mathrm{A}$ of 39$\pm$12 mJy found for the H76$\alpha$ line is a factor $\sim$2 lower than that of 114 mJy as measured by \cite{Goss85} for the entire region A, which agrees with the fact that only half of the region A falls inside the GBT beam size toward \emph{LOS}$-$0.02. Despite this finding, we believe that it is unlikely that the GBT data traces extended RRL emission toward \emph{LOS}+0.693 and \emph{LOS}$-$0.11 and that it does not trace extended RRL emission toward \emph{LOS}$-$0.02.
In fact this is supported in Fig.~\ref{CII_RRL_spectra} (upper panel) where we show the C{\small \,I}, C{\small \,II} and H79$\alpha$ spectra of \emph{LOS}$-$0.11 and \emph{LOS}$-$0.02. It can be seen that both the positive and negative velocity components of the extended ionised gas traced by the C{\small \,II} emission \citep{Garcia15} are also well traced by the H79$\alpha$ line emission. Therefore, in this paper we consider that the GBT traces extended ionised gas in the three \emph{LOSs}.

The studied ionised gas is also diffuse because we have found n$_\mathrm{e}$ of $\sim$40$-$310 cm$^{-3}$, which are much lower than those of 3600$-$5100 cm$^{-3}$ found in compact H{\small \,II} regions of the GC \citep{Mills11}. The n$_\mathrm{e}$ of $\sim$40$-$120 cm$^{-3}$ found for the negative velocity gas of \emph{LOS}$-$0.11 and \emph{LOS}$-$0.02 are consistent with those of $\sim$100$-$130 cm$^{-3}$ found for diffuse ionised gas of the Arched Filaments H{\small \,II} complex \citep{Langer17}. 

C{\small \,II} emission traces a negative velocity component not only in both Sgr A sources but also in the H1 source and Sgr A* (see the bottom panel of Fig.~\ref{CII_RRL_spectra}). We also note in Fig.~\ref{CII_RRL_spectra} that the C{\small \,I} emission does not trace the negative velocity component in \emph{LOS}$-$0.11 and \emph{LOS}$-$0.02 but it does partially in the H1 source and Sgr A*.

\subsection{Kinematics of the ionised gas}\label{Kinematics}

\begin{figure}
\includegraphics[width=0.5\textwidth]{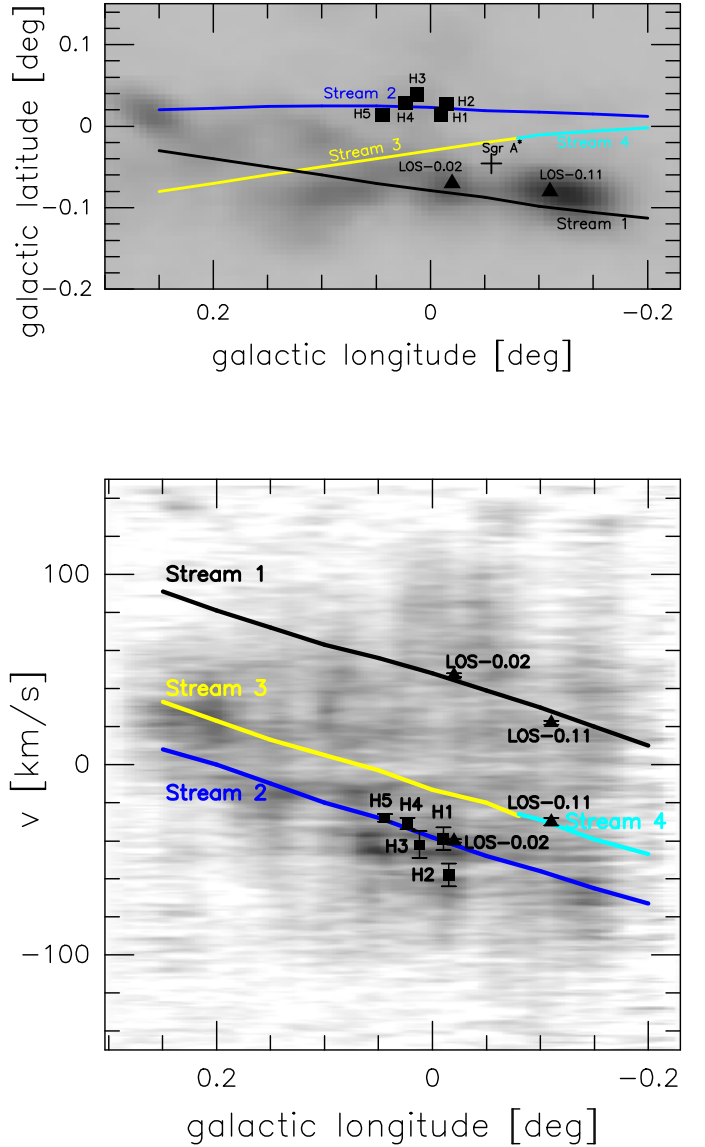}
\caption{\textbf{Upper panel:} Four streams used to model the GC gas kinematics \citep{Kruijssen15}. The NH$_3$(1,1) emission map obtained using HOPS data \citep{Purcell12} is shown in gray scale. The map is integrated over the velocity range -100 to +100 km s$^{-1}$. Filled triangles indicate positions of \emph{LOS}$-$0.11 and \emph{LOS}$-$0.02, while filled squares show the positions of H1$-$H5 sources. The black cross shows the position of Sgr A*. \textbf{Bottom panel:} The four streams are drawn on the position-velocity C{\small \,II} map obtained using HIFI data \citep{Garcia16}. This map covers $0.1\degr$ in latitude centered at $0\degr$. Central line velocities of \emph{LOS}$-$0.11 and \emph{LOS}$-$0.02 are shown with filled triangles, while the central line velocities of H1$-$H5 sources \citep{Zhao93} are indicated with filled squares. Velocity error bars overlap with filled triangles.}\label{pos_vel}
\end{figure}

Our RRLs show diffuse ionised gas with negative and positive velocities in \emph{LOS}$-$0.11 and
\emph{LOS}$-$0.02.
In Fig.~\ref{pos_vel} we show the velocities of both Sgr A sources on a position-velocity diagram for the C{\small \,II} emission. Four gas streams from the model of \cite{Kruijssen15} are also shown in this figure. The $v_r$ of both velocity components of \emph{LOS}$-$0.11 are consistent with the velocities of streams 1 and 4, while the $v_r$ of both velocity components of \emph{LOS}$-$0.02 agree with the velocities of streams 1 and 2. This suggests that along the two \emph{LOSs} the GBT traces diffuse and extended ionised gas that are part of the gas streams orbiting the GC. The positions of \emph{LOS}$-$0.11 and \emph{LOS}$-$0.02 (see Fig.~\ref{pos_vel}, upper panel) also support that the positive velocity gas in both sources is part of stream 1.
The velocities and positions of the regions H1$-$H5 (see Fig.~\ref{pos_vel}) show that these sources are likely associated with stream 2. If this hypothesis is correct then the negative velocity gas of \emph{LOS}$-$0.02 could coexist with the H1$-$H5 sources.
Our hypothesis is in agreement with the finding of \cite{Langer17} that the kinematics of the ionised gas in the Sgr A and Sgr B2 complexes, as traced by the C{\small \,II} emission, is well explained by the gas streams proposed by \cite{Kruijssen15}.

\subsection{Sources of ionisation}\label{Ionization}

\subsubsection{Positive velocity gas in the three GC \emph{LOSs}}

The ionised gas studied in \emph{LOS}$-$0.11 is at a projected distance of \mbox{$\sim$3.8 \textrm{pc}} from the region G (see Fig.~\ref{fig1}), whose massive stars are thought to be the closest compact source of ionisation to the positive velocity \emph{LOS}$-$0.11 gas \citep{Ho85}. On the other hand, since the region L lies close to \emph{LOS}+0.693 (see Fig.~\ref{fig1}, bottom panel), it is expected that the main ionisation source of the \emph{LOS}+0.693 gas could be the massive stars responsible for the ionisation of the \mbox{region L}.

As can be seen in the upper panel of Fig.~\ref{fig1}, part of the emission arising in the region A falls within the GBT beam toward \emph{LOS}$-$0.02. Thus we expect that the gas with positive velocities in this GC source is mainly ionised by the massive stars in the H{\small \,II} region A.
To test whether the H{\small \,II} region A could be the main source of ionisation of the positive velocity gas in \emph{LOS}$-$0.02
we estimate the number of photons inside the GBT beam, $N_{\Omega}$, and that $\sim$50\% of the region A falls inside the GBT beam (see Fig.~\ref{fig1}). Considering the location of the GBT beam centre, then the compact H{\small \,II} region A would actually be displaced from the telescope beam centre. For this geometry we can estimate an upper limit to $N_{\Omega}$ following the expression given by \cite{Rodriguez05}:

\begin{equation}\label{equa3}
 N_{\Omega}=\frac{N_\mathrm{Lyc}}{4\pi r^2} \Omega D^2 \mathrm{,}
\end{equation}

\noindent where $r$ is the radius of the H{\small \,II} region. We derive the upper limit of 10$^{50.95}$ photons s$^{-1}$ for the value of $N_{\Omega}$ by using the $N_\mathrm{Lyc}$ value provided by \cite{Mills11} for the H{\small \,II} region A, $\Omega$=45 arcsec and $r$=1.7 pc in \mbox{Eq. \ref{equa3}.}
It seems that the positive velocity gas of \emph{LOS}$-$0.02 is mainly ionised by the massive stars in the 
H{\small \,II} region A because the 
$N_\mathrm{Lyc}$ values of \emph{LOS}$-$0.02, given in Table \ref{table9}, are consistent with the upper limit of 10$^{50.95}$ photons s$^{-1}$.

\subsubsection{Negative velocity gas in \emph{LOS}$-$0.11 and \emph{LOS}$-$0.02}
 
The ionised gas components with negative velocities found toward both sources of \mbox{Sgr A} raises the question of the source of ionisation. The top$-$down view shown in Fig. 6 of \cite{Kruijssen15} gives us information about the distances between Sgr A* and the four streams considered in their kinematical model. In this scenario Sgr A* is located between both the 20 and 50 km s$^{-1}$ clouds and their background gas streams 3 and 4, at a projected distance of $\sim$60 pc from these features.
If the negative velocity \emph{LOS}$-$0.02 gas is part of the gas stream 2, as discussed in \mbox{Section \ref{Kinematics},} 
then it may be ionised by the photons arising in massive O6$-$O7 stars which also ionise the presumably closest UC$-$H{\small \,II} regions, \textrm{i.e.} H1$-$H5 \citep{Zhao93}, located at least \mbox{$\sim$12 \textrm{pc}} away from the negative velocity gas observed 
toward \emph{LOS}$-$0.02 (see Fig.~\ref{pos_vel}, upper panel). 
Of course, other ionising sources apart from those proposed may exist in the environment of the negative velocity \emph{LOS}$-$0.11 gas. On the other hand, the negative velocity \emph{LOS}$-$0.02 gas is likely part of the stream 4, as discussed in Section \ref{Kinematics}.
So far there are no compact H{\small \,II} regions or massive stars whose velocities and positions are consistent with those of the gas stream 4 around \emph{LOS}$-$0.11, hence the identification of ionising sources of the negative velocity \emph{LOS}$-$0.11 gas remains unclear. A possibility is that the negative velocity \emph{LOS}$-$0.11 gas is actually part of stream 2, despite the difference in their velocites (see Fig.~\ref{pos_vel}, bottom panel), thus being also ionised by the massive stars inside H1$-$H5 sources as for \emph{LOS}$-$0.02.
 
Considering the gas stream model proposed by \cite{Kruijssen15} the massive young stars orbiting Sgr A* can be ruled out as ionising sources of the negative velocity \emph{LOS}$-$0.11 and \emph{LOS}$-$0.02 gas since in this scenario Sgr A* is $\sim$60 pc away from gas streams 2 and 4 along the two \emph{LOSs}.

\section{Conclusions}\label{Conclusions}

Using the GBT telescope we have detected extended and diffuse ionised emission toward three GC \emph{LOSs}. The main conclusions of the present work are as follows:

\begin{itemize}
\item We found that the ionised gas observed toward the three GC sources is emitted under LTE conditions based on the H$m\beta$$-$to$-$H$n\alpha$ integrated line intensity ratios.
\item We found a $^4$He mass fraction Y of \mbox{0.29$\pm$0.01} that supports the hypothesis that high-mass stars in the GC have enriched the helium$-$4 abundance in the ISM as compared to the primordial value.
\item For \emph{LOS}$-$0.11, \emph{LOS}$-$0.02 and \emph{LOS}+0.693 we have derived $n_\mathrm{e}$ and $N_\mathrm{Lyc}$ values. The studied gas is characterised by $n_\mathrm{e}$ of $\sim$40$-$310 cm$^{-3}$.
\item The ionised gas detected toward regions of the 20 and 50 km s$^{-1}$ clouds is likely associated, following the \cite{Kruijssen15} model, with gas stream 1 orbiting the GC, while the ionised gas moving with negative velocities in \emph{LOS}$-$0.02 and \emph{LOS}$-$0.11 is likely associated with the gas streams 2 and 4, respectively, located in projection $\sim$12 pc above stream 1.
\item The \emph{LOS}$-$0.02 gas at positive velocities is mainly ionised by UV photons produced in the massive stars also ionising the H{\small \,II} region A. The massive stars inside the H{\small \,II} regions L and G are considered the closest sources of gas ionisation of \emph{LOS}+0.693 and \emph{LOS}$-$0.11 (positive velocity component), respectively.
\item We propose that the gas with negative velocities observed toward \emph{LOS}$-$0.02 may be ionised by UV photons originating in the massive stars of the presumably closest H{\small \,II} regions H1$-$H5.
\item The negative velocity gas observed toward \emph{LOS}$-$0.11 is likely associated with gas stream 4. We were not able to propose any possible ionising sources of the negative velocity \emph{LOS}$-$0.11 gas because, so far, there are no compact H{\small \,II} regions or massive stars having both velocities and positions similar to those expected for gas stream 4 around \emph{LOS}$-$0.11. However, if the negative velocity components of both Sgr A sources are part of the stream 2, then the massive stars in the H1$-$H5 regions could be the main sources of UV photons ionising the gas with negative velocities of both Sgr A sources. 
\item We compared C{\small \,I} spectra with our H79$\alpha$ spectra, finding that C{\small \,I} emission does not trace the negative velocity component of either of the Sgr A sources. This indicates that this diffuse gas component is fully ionised.

\end{itemize}
\section*{Acknowledgements}
We thank the anonymous referee for comments which helped to improve this paper.
A. B\'aez-Rubio acknowledges support from a DGAPA postdoctoral grant (year 2015) to UNAM. J.M.-P. acknowledges partial support by the MINECO under grants ESP2015$-$65597$-$C4$-$1 and ESP2017$-$ and Comunidad de Madrid grant number S2013/ICE$-$2822 SpaceTec$-$CM. 











\bsp	
\label{lastpage}
\end{document}